\documentclass[preprint2]{aastex}
\usepackage{natbib}

\usepackage{color}
\definecolor{blue}{rgb}{0.1,0.1,0.6}
\definecolor{orange}{rgb}{0.74,.35,0.099}
\definecolor{pale}{rgb}{0.90,0.90,0.95}
\definecolor{red}{rgb}{1.0,0.0,0.0}

\newcommand{\degree}{\ensuremath{^\circ}}

\shorttitle{GPI Spectroscopy of HR 8799 c and d}
\shortauthors{Ingraham et al.}

\begin{document}


\title{Gemini Planet Imager Spectroscopy \break of the HR 8799 planets c and d}


\author{Patrick Ingraham\altaffilmark{1}, 
Mark S. Marley \altaffilmark{2}, 
Didier Saumon \altaffilmark{3}, 
Christian Marois \altaffilmark{4}, 
Bruce Macintosh \altaffilmark{1}, 
{Travis Barman}\altaffilmark{5}, 
{Brian Bauman}\altaffilmark{6}, 
{Adam Burrows}\altaffilmark{7}, 
{Jeffrey K. Chilcote}\altaffilmark{8}, 
{Robert J. De Rosa}\altaffilmark{9,10}, 
{Daren Dillon}\altaffilmark{11}, 
{Ren\'e Doyon}\altaffilmark{12}, 
{Jennifer Dunn}\altaffilmark{4}, 
{Darren Erikson}\altaffilmark{4}, 
{Michael P. Fitzgerald}\altaffilmark{8}, 
{Donald Gavel}\altaffilmark{11}, 
{Stephen J. Goodsell}\altaffilmark{13}, 
{James R. Graham}\altaffilmark{14}, 
{Markus Hartung}\altaffilmark{13}, 
{Pascale Hibon}\altaffilmark{13}, 
{Paul G. Kalas}\altaffilmark{14}, 
{Quinn Konopacky}\altaffilmark{15}, 
{James A. Larkin}\altaffilmark{8}, 
{J\'{e}r\^{o}me Maire}\altaffilmark{15}, 
{Franck Marchis}\altaffilmark{16}, 
{James McBride}\altaffilmark{14}, 
{Max Millar-Blanchaer}\altaffilmark{15}, 
{Katie M. Morzinski}\altaffilmark{17}, 
{Andrew Norton}\altaffilmark{11}, 
{Rebecca Oppenheimer}\altaffilmark{18}, %
{Dave W. Palmer}\altaffilmark{6}, 
{Jenny Patience}\altaffilmark{9}, 
{Marshall D. Perrin} \altaffilmark{19}, 
{Lisa A. Poyneer}\altaffilmark{6}, 
{Laurent Pueyo}\altaffilmark{19},  
{Fredrik Rantakyr\"o}\altaffilmark{13}, 
{Naru Sadakuni}\altaffilmark{13}, 
{Leslie Saddlemyer}\altaffilmark{4}, 
{Dmitry Savransky}\altaffilmark{20}, 
{R\'emi Soummer}\altaffilmark{19}, 
{Anand Sivaramakrishnan}\altaffilmark{19}, 
{Inseok Song}\altaffilmark{21}, 
{Sandrine Thomas}\altaffilmark{2,22}, 
{J. Kent Wallace}\altaffilmark{23}, 
{Sloane J. Wiktorowicz}\altaffilmark{11}, 
{Schuyler G. Wolff}\altaffilmark{19}, 
}

\altaffiltext{1}{Kavli Institute for Particle Astrophysics and Cosmology, Stanford University, Stanford, CA 94305, USA}
\altaffiltext{2}{NASA Ames Research Center,  Moffett Field, CA 94035, USA}
\altaffiltext{3}{Los Alamos National Laboratory, Los Alamos, NM 87545, USA}
\altaffiltext{4}{NRC Herzberg Astronomy and Astrophysics, 5071 West Saanich Road, Victoria, BC V9E 2E7, Canada}
\altaffiltext{5}{Lunar and Planetary Laboratory, University of Arizona, Tucson, Arizona 85721-0092, USA}
\altaffiltext{6}{Lawrence Livermore National Lab, 7000 East Ave., Livermore, CA 94551, USA}
\altaffiltext{7}{Department of Astrophysical Sciences, Princeton University, Princeton, NJ 08544, USA}
\altaffiltext{8}{Department of Physics and Astronomy, UCLA, Los Angeles, CA 90095, USA}
\altaffiltext{9}{School of Earth and Space Exploration, Arizona State University, PO Box 871404, Tempe, AZ 85287, USA}
\altaffiltext{10}{School of Physics, College of Engineering, Mathematics and Physical Sciences, University of Exeter, Stocker Road, Exeter, EX4 4QL, UK}
\altaffiltext{11}{Department of Astronomy, UC Santa Cruz, 1156 High Street, Santa Cruz, CA 95064, USA}
\altaffiltext{12}{Department de Physique, Universit\'{e} de Montr{\'e}al, Montr\'eal QC H3C 3J7, Canada}
\altaffiltext{13}{Gemini Observatory, Casilla 603, La Serena, Chile}
\altaffiltext{14}{Department of Astronomy, UC Berkeley, Berkeley CA, 94720, USA}
\altaffiltext{15}{Dunlap Institute for Astrophysics, University of Toronto, 50 St. George St, Toronto ON M5S 3H4, Canada}
\altaffiltext{16}{SETI Institute, Carl Sagan Center, 189 Bernardo Avenue, Mountain View, CA 94043, USA}
\altaffiltext{17}{NASA Sagan Fellow, Steward Observatory, University of Arizona, 933 N. Cherry Ave., Tucson, AZ 85721, USA}
\altaffiltext{18}{American Museum of Natural History, New York, NY 10024, USA}
\altaffiltext{19}{Space Telescope Science Institute, 3700 San Marting Drive, Baltimore, MD 21218, USA}
\altaffiltext{20}{Sibley School of Mechanical and Aerospace Engineering, Cornell University, Ithaca, NY 14853, USA}
\altaffiltext{21}{University of Georgia, Department of Physics and Astronomy, Athens, GA 30602, USA}
\altaffiltext{22}{UARC, UC Santa Cruz, Santa Cruz, CA 95064, USA}
\altaffiltext{23}{Jet Propulsion Laboratory/California Institute of Technology, 4800 Oak Grove Dr., Pasadena, CA 91109, USA}


\begin{abstract}
During the first-light run of the Gemini Planet Imager (GPI) we obtained $K$-band spectra of exoplanets HR 8799 c and d. Analysis of the spectra indicates that planet d may be warmer than planet c. Comparisons to recent patchy cloud models and previously obtained observations over multiple wavelengths confirm that thick clouds combined with horizontal variation in the cloud cover generally reproduce the planets' spectral energy distributions. When combined with the 3 to $4\, \, \mu m$ photometric data points, the observations provide strong constraints on the atmospheric methane content for both planets. The data also provide further evidence that future modeling efforts must include cloud opacity, possibly including cloud holes, disequilibrium chemistry, and super-solar metallicity.

\end{abstract}


\section{Introduction}

To date, our current understanding of the atmospheres of directly imaged exoplanets is primarily (but not exclusively) developed from fitting their broadband photometric colors and studying the spectra of free-floating planetary mass objects ($m<13 \ M_J$) and brown dwarfs (75$\ M_J>m>13 \ M_J$). A new suite of dedicated high-contrast imaging spectrometers now being commissioned \citep[GPI, SPHERE, ScEXAO,][]{Macintosh14a, Beuzit08a,Martinache09a} will enable the study of exoplanets beyond 5 AU from their host stars, a parameter space largely unexplored via indirect techniques. Observations from these instruments will probe the atmospheres via spectroscopy and constrain the orbital parameters through astrometric measurements.

The HR 8799 system contains four known planetary-mass companions, all discovered at near-infrared wavelengths \citep[][]{marois08a,marois10a}. Atmospheric models for these planets have struggled to reproduce the near-infrared colors while still reporting realistic radii \citep[][]{marois08a,Bowler10a, Currie11a}. The inclusion of broadband photometric data points between 3-5 $\mu$m \citep[][]{Hinz10a, Janson10a, Galicher11a, Skemer12a}, where these planets release a significant fraction of their energy, imply that cloudy atmospheres combined with effects of non-equilibrium chemistry are necessary to adequately reproduce the colors of these $\sim$1000 K, low-gravity objects \citep[][]{Madhusudan11a, Marley12a}.

Obtaining spectra of the HR 8799 planets is extremely challenging. Spectroscopy of HR 8799 b, c, d, and e in $J$ and $H$-bands was carried out using the high-contrast imaging instrumentation Project 1640 \citep[][]{Hinkley11a,Oppenheimer13a} at a relatively low signal-to-noise ratio (SNR). To date, only planets b and c have spectroscopic measurements in $K$ band \citep[][]{Barman11a, Konopacky13a}, both using the OSIRIS instrument on the Keck telescope, a multi-purpose medium-resolution ($\lambda/\Delta \lambda \sim 4000$) integral field spectrograph \cite[][]{Larkin06a}. 

$K$-band emergent spectra of young giant planets are sensitive to atmospheric thermal structure, gas composition ($\rm H_2$ pressure induced absorption, $\rm H_2O$, CO, and $\rm CH_4$ are the most notable opacity sources), and cloud opacity. 

Perhaps the greatest surprise arising from spectra of the HR 8799 b, c, and d planets has been the notable lack of atmospheric methane detected in $H$ and $K$ bands \citep[][]{Bowler10a, Barman11a, Currie11a, Konopacky13a}. Field brown dwarfs with comparable effective temperatures ($T_{\rm eff}$) show strong methane signatures at these wavelengths. The lack of methane has been attributed to a combination of atmospheric mixing, driving the gas composition away from equilibrium, as well as heating caused by thick silicate clouds \citep[][]{Barman11a,Currie11a,Marley12a}. 

In this letter, we present $K$-band spectroscopic observations of HR 8799 c, and for the first time, planet d, obtained during first-light observations of the Gemini Planet Imager \citep[GPI,][]{Macintosh14a, Macintosh14b, Larkin14a}; a newly commissioned facility instrument dedicated to the detection and characterization of extrasolar planets. We compare the spectra to current models \citep[][]{Marley12a} optimized to fit the photometric data points of several bandpasses. Lastly, the implications of the spectra for future modeling efforts are discussed.

\section{Observations and Data Reduction}

During the first commissioning run HR 8799 was observed in the \emph{K1} spectral band (1.90-2.19 $\mu$m) on November 17, 2013 at an airmass of $\sim$1.65 and a median integrated seeing of 0\farcs97 (at 500 nm) as measured at zenith by the Gemini Differential Image Motion Monitor. These conditions are significantly worse than GPI's standard operating design specification. However, moderate SNR spectra of planets c and d were still obtained. Planet b falls outside the field of view. A total of sixteen 90-second exposures were obtained; the last ten images had the IFS cryocoolers power decreased to 30\% to minimize vibration in hopes of a moderate increase in AO performance. Due to the high airmass and poor seeing contrast was dominated by residual atmospheric turbulence rather than vibrations, therefore all images were weighted equally during the image reduction to minimize the photon noise and maximize the field rotation ($\sim$9.1$\degree$). Four 90-second sky images were taken directly after the observations. 



The following night observations were performed in the \emph{K2} spectral band (2.12-2.38 $\mu$m) in better seeing conditions (0\farcs75) but still at high airmass ($\sim$1.63). A total of twenty 90-second exposures covered $\sim$10.5$\degree$ of field rotation. The cryocooler's power was decreased for all but the first 6 frames. Five 90-second sky frames were taken directly afterwards. During the first light run, the atmospheric dispersion corrector was not yet available for use. This necessitated that the individual wavelength slices of each datacube be registered to a common position in post-processing prior to invoking any speckle suppression algorithms.

The data reduction was accomplished using the GPI Data Reduction Pipeline (DRP) \citep[][]{Perrin14a}. The wavelength calibration was determined using the xenon arclamp. However, the zero-point flexure offset between the calibration and observation position was determined by fitting the atmospheric transmission spectrum multiplied by our instrument response function to the telluric absorption lines (in the \emph{K1} spectrum) and the filter edges (in the \emph{K2} spectrum). Comparison of this method with known offsets has proved it to be consistent to within $\sim$0.1 pixels ($\sim$10 nm). A master dark and median sky spectrum was subtracted from the raw images and the bad pixels corrected. The individual raw images were then transformed into 3-dimensional spectral datacubes. To compensate for the atmospheric refraction the individual wavelength slices were registered prior to performing point-spread function (PSF) subtraction using the TLOCI technique \citep[][]{Marois14a}. For this reduction, TLOCI utilized only angular differential imaging techniques and no spectral differential imaging to ensure a proper spectral extraction. The \emph{K1} and \emph{K2} spectra were merged using the overlapping regions and then normalized using the photometry from \citet[][]{marois08a} in the region of common filter bandpasses.

The atmospheric and telescope+instrument transmission functions were removed using measurements of the host star spectrum, obtained by extracting the spectra of the four fiducial images of the host star that are present in every GPI coronagraphic image \citep[][]{Wang14a}. The error bars for the companion's spectra at each wavelength are a combination of the standard deviation of the background, convolved with the PSF model in the datacubes and the four fiducial stellar spectra. For planets c and d the surrounding background noise in the TLOCI reduced datacube at the same angular separation appears Gaussian in nature, indicating that residual speckle noise is not the dominant error term and the uncertainties are indeed properly characterized by a 1-sigma error bar. The leading causes of uncertainty are the variation of the stellar spectrum determined from each satellite spot, the speckle photon noise, and thermal/sky background noise. 

\section{Results}

\subsection{HR 8799c}

The first $K$-band spectrum of HR 8799 c was obtained by \citet[][]{Konopacky13a} who performed observations for 5.5 hours using the Keck telescope and OSIRIS instrument. Although their spectrum was obtained at a significantly higher spectral resolution, the error bars on their spectrum when binned to GPI's resolution are comparable to what GPI obtained in $\sim$30 minutes in both \emph{K1} and \emph{K2}, in relatively poor observing conditions. 

\begin{figure}[ht]
\begin{center}
\includegraphics[scale=0.36]{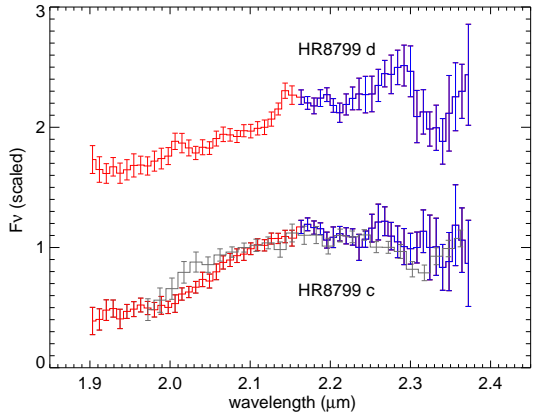} 
\caption{ \label{fig:cdspec} Lower spectrum shows HR 8799 c with the \emph{K1} and \emph{K2} bands shown in red and blue respectively. The black dotted line with grey error bars indicate the spectrum from \citet[][]{Konopacky13a}. The spectrum of HR 8799 d, shown above c, has been offset by 1.0 for clarity.  (A color version of this figure is available in the online journal.) }
\end{center}
\end{figure}

Figure \ref{fig:cdspec} shows the consistency of both the GPI and \citet[][]{Konopacky13a} spectra. The presence of the CO band at 2.295 $\mu$m is not detectable due to our large error bars in this region. The increased errors redder than $\sim$2.3 $\mu$m is a result of a decrease in the instrument transmission and an increase in the thermal background. The small differences between the 2.01-2.06 $\mu$m regions is most likely due to the presence of strong telluric $\rm CO_2$ absorption lines that are not being adequately treated in the spectrum. 

\subsection{HR 8799d}

The upper spectrum in Figure \ref{fig:cdspec} shows the first $K$-band spectrum obtained of HR 8799d. As discussed below, the broad shape is significantly flatter than planet c indicating that previous predictions of HR 8799 c and d having similar mass, composition, cloud properties, and effective temperature may be incorrect.  

One notable feature is an increase in flux between $\sim$2.25 and $\sim$2.29 $\mu$m. The width of this feature is consistent with the width of a speckle moving through the companion PSF. However, a histogram of the pixel values in an empty region surrounding the object suggests that the SNR of the extracted spectrum is not speckle noise limited. Errors introduced by the correction of the stellar spectrum via the satellite spots may introduce systematics not adequately characterized by a Gaussian. However, we know the $\sim$2.285 $\mu$m bump is not introduced by the satellite spots because it is not present in the spectrum of HR 8799 c.  

The only plausible molecular compounds we have found with opacity features in this spectral region are $\rm C_2H_2$ and HCN, either of which would have to be present in emission. HCN is the third most abundant carbon molecule, formed as an intermediate product of the disequilibrium CO to $\rm CH_4$ conversion, while $\rm C_2H_2$ is a product of methane photochemistry. Both molecules, however, have stronger features near 2.11 and particularly $3\,\rm \mu m$ that are not apparent in the data. 
The presence of the CO $\Delta v$ = 2-0 band head at 2.295 $\mu$m appears strong but in fact is not statistically significant due to our uncertainties in the region. 

\section{Discussion}

\subsection{Comparison to previous work}

\begin{figure*}[ht]
\begin{center}
\includegraphics[scale=0.6,angle=270,origin=c, trim=6cm 4cm 6cm 6cm]{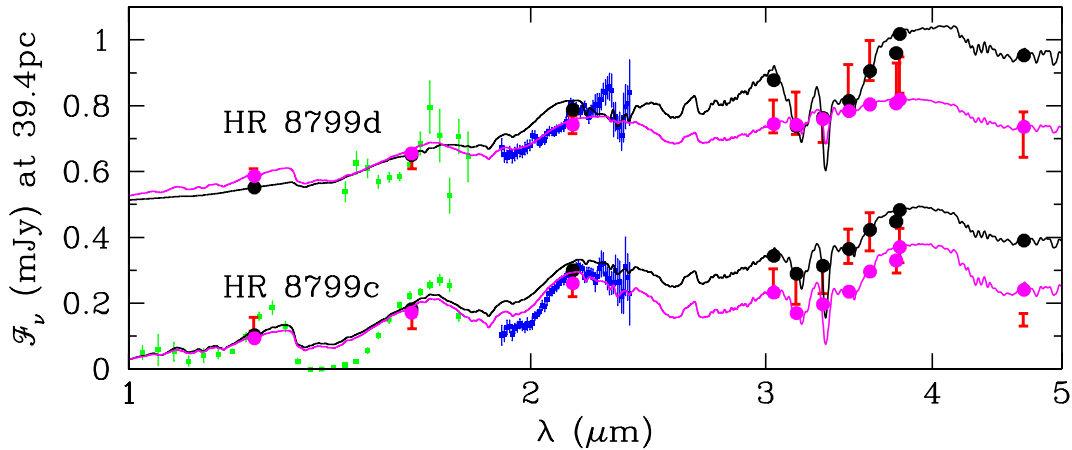} 
\caption{ \label{fig:modeldata} A comparison of synthetic spectra with the data for planets c and d. The fluxes for planet d have been shifted upward by 0.5$\,$mJy for clarity. Photometric data compiled by \citet[][]{Skemer14a} are shown as red error bars. Project 1640 spectra \citep[][]{Oppenheimer13a} are shown in green, and the GPI spectra in blue.  The black curves are cloudy models with $T_{\rm eff}=1100\,$K, $g=10^4\,$cm s$^{-2}$ and solar abundances.  The model for planet c has a cloud sedimentation parameter $f_{\rm sed}=0.25$ and a partial cloud cover with 5\% holes (see text).  The model for planet d has $f_{\rm sed}=0.5$ and no holes. Black dots show the corresponding synthetic photometry. These model fluxes are scaled with the radius from the evolution models \citep{sm08}. The magenta lines indicate the best fitting models (discussed in the text) to the GPI $K$-band data alone, where the radii are normalized to produce the observed $K$-band flux. (A color version of this figure is available in the online journal.)}
\end{center}
\end{figure*}

Previous studies comparing models to photometry have tended to assign HR 8799 c and d similar masses and effective temperature (summarized in Table 1 of \citealt[][]{Marley12a}) but the $K$-band spectra (Figure \ref{fig:cdspec}) illustrate that there are significant differences between the planets. The d spectrum is flatter with a somewhat shallower water absorption band to the blue. Figure \ref{fig:modeldata} places the GPI data for both planets in context with the photometry tabulated in \citet[][]{Skemer12a, Skemer14a} (red) and the spectra of \citet[][]{Oppenheimer13a} (green). The latter are normalized to the $H$-band flux and the GPI spectra (blue) to the $K_s$ flux.

\subsection{Fitting evolutionary and atmosphere models}

To understand which classes of model spectra are most similar to this combined dataset we first computed a range of atmosphere models with varying gravity, effective temperature, cloud properties, and strength of eddy mixing with planet radii computed from `hot start' evolution models \citep{sm08}. The models, all of which utilize solar abundances \citep{lodders2003}--including for C and O ($\rm C/O = 0.50$ vs.\ 0.55 in \citet{asplund2009})--are described in more detail in \citet[][]{Skemer14a}. Figure \ref{fig:modeldata} compares two of these models (black curves) to data.  As others have found (e.g. \citet[][]{Currie11a,Barman11a,Marley12a}), no single model fits all of the available broadband data. Most challenging is the need to reproduce the spectral peaks in the $J$ band, which implies relatively long, unobscured gas columns allowing flux to emerge from the deep atmosphere, while keeping the atmospheric abundance of methane at undetectable levels in the $H$ and $K$ bands. The relatively cloudless column needed to match $J$-band photometry tends to promote cooling and thus the formation of methane. Previous models have depended on particularly thick clouds to keep the atmosphere warm and methane-free, but too much cloud opacity shuts off the $J$ window and renders the atmospheric emission nearly Planckian. Allowing for non-equilibrium chemistry, namely the mixing of deep CO to the observable atmosphere, helps reproduce the absence of methane absorption, but simultaneously fitting the $L^\prime$ photometry of \citet[][]{Skemer14a}, the $M$-band photometry \citep[][]{Galicher11a}, and the GPI $K$-band spectra is difficult. 

Despite considering a broad range of solar-composition, cloudy and partially cloudy atmosphere models we did not find a single fully satisfactory set of models that adequately reproduce all available data. To illustrate this conundrum, we describe here two modeling assumptions which imply somewhat different planetary parameters.
The first class of models utilizes planetary radii as a function of $T_{\rm eff}$ and $\log g$ as predicted by standard evolution models, while in the second case (described in the next sub-section) we allow the radius to vary independently of the computed evolution. Given the resulting uncertainty in our understanding we 
provide neither formal best fits nor error ranges on estimated model parameters.

In the first case, the best fits to  the data for planets c and d are obtained for models with thick clouds, which raise the atmospheric temperature and helps to favor CO, combined with some horizontal variation in cloud cover (Figure 2, black lines). Allowing for small cloud-free patches (following \citealt[][]{Marley10a}, see also \citealt[][]{Currie11a}) permits more flux to emerge in the $J$ band than a fully cloudy case while not cooling the overall column excessively. Specifically we find $T_{\rm eff} \sim 1100\,\rm K$, $\log g = 4$, $f_{\rm sed}$ (our cloud sedimentation parameter \citealt[][]{Ackerman01a}) in the range of 0.25 to 0.50, 5 to 10\% cloud holes, and employing the `hot-start' evolution radius produces model spectra that are qualitatively consistent with much of the data for both planets. 

The models shown in black in Figure \ref{fig:modeldata}, however, are certainly not ideal. They over-predict the $M$-band flux for both planets. Further increasing the efficiency of vertical mixing does not help as nearly all the carbon is already locked in CO for eddy diffusion coefficient $K_{zz}=10^4\,\rm cm^{2}\,s^{-1}$ (see \citet{Skemer14a} for further explanation). Excess $\rm CO_2$ and perhaps PH$_3$ over that predicted by the model might account for the remaining discrepancies. Higher than solar metallicity, as suggested by \citet[][]{Barman11a} for planet b, could also be responsible. Additionally these models do not fit the GPI $K$-band spectra particularly well.

For HR 8799c none of the models we explored were able to reproduce the broad, deep water absorption band at $1.4\,\rm\mu m$ seen in the Project 1640 data \citep[][]{Oppenheimer13a} while still maintaining consistency with the water band depth seen in the GPI data at $1.9\,\rm \mu m$. The modeled relative depth of these two water bands is, however, generally consistent with trends seen in the spectra of late L-dwarfs \citep[][]{Cushing05a, Stephens09a}.  

\subsection{Fitting normalized atmosphere models}

As noted in the discovery paper \citep{marois08a} and summarized in Table 1 of \citet[][]{Marley12a}, better fits to the observations of some of the HR 8799 planets can be obtained by relaxing the requirement that the planet radius for a specified $T_{\rm eff}$ and $\log g$ be consistent with that expected from evolution models. Thus for our second modeling approach we fit spectra only to the GPI $K$-band data for 8799 c and d. For each model we simply normalized the model spectrum to the observations and identified which model spectrum best fit the data in a least squares sense. We then chose a model radius that would account for the necessary flux normalization and compared the resulting spectrum to all of the available data. The results of this exercise are shown in Figure \ref{fig:modeldata} (magenta lines).

For planet d the model that best fits the $K$-band data was for $T_{\rm eff}=1300\,\rm K$, $\log g = 4$, $f_{\rm sed}=0.5$ and with 5\% spatial coverage of cloud opacity `holes' that had 1\% of the background cloud opacity. The normalized radius is 59,000 km ($0.84\,\rm R_J$), or about 56\% of the evolution radius for these parameters of 106,000 km ($1.52\,\rm R_J$). Although this model was only fit to $K$-band, the agreement with most of the remaining photometry is quite good. The $M$-band photometric point that the other models could not fit is notably well matched. Taken at face value, such a radius implies a planet mass of $\sim$3$\,\rm M_J$ for $\log g = 4$. However, a solar composition gas giant planet with this mass would have a radius in excess of 100,000\,$\rm km\, (1.43\,R_J)$ for standard evolution model assumptions \citep[][]{FMB07}. A planet with a massive core, equal to about 60 to 70\% the mass of the planet, with a solar composition $\rm H_2$-He envelope would have the appropriate radius \citep{FMB07,bcb08}. However evolution models have never been computed for such a planet and it is unclear if such a high $T_{\rm eff}$ could be maintained for as long as the age of the system.

For planet c the best $K$-band fits, driven by the spectral slope to the blue side of the band, also favor lower gravity models, although the gravity is poorly constrained. Figure 2 illustrates an example with
$T_{\rm eff}=1300\,\rm K$, $\log g = 3.75$, $f_{\rm sed}=1$. This is a slightly lower gravity and less cloudy model than the best fit for planet d, but with no cloud holes. With the same radius scaling method as planet d, the $K$-band photometric point is matched but the overall fit to the other available data is not as good. A planet with this gravity and such a scaling from the evolution radius would have a radius of $63,200\, \rm km$ ($0.90\,\rm R_J$) and a mass of $\sim$2$\,\rm M_J$. Again only a planet with a heavy element fraction in excess of 50\% of the total mass would match such parameters although we stress that both the mass and radius are poorly constrained by the fit. 

We regard these small radii solutions as unlikely and do not explore them in detail as they imply exceptionally massive cores. Furthermore, a new class of evolution models for such planets are required to evaluate the plausibility of such a combination of parameters at this age. We expect that the most likely solution would be models that combine many of the characteristics that have been previously recognized: thick clouds with some holes, vertical mixing, enhanced heavy element abundances leading to smaller radii (but not as extreme as these examples), and higher metallicity atmospheres, such as already inferred for HR 8799b \citep{Lee13a}. 

\section{Conclusion}

As noted in \citet[][]{Skemer14a}, the 3 to $4\, \, \mu m$ photometry suggests the presence of a small amount of $\rm CH_4$ in the atmosphere of planet d. This provides a rather strong constraint on the models as the methane column must be large enough to be detectable at $3.3\,\rm \mu m$, but not in the GPI or Keck $K$ band data. The now well established trend of methane deficiency at low gravity \citep{Barman11b} not only points to disequilibrium chemistry but also to a shortcoming in current models of CO to $\rm CH_4$ conversion. New work on CO to methane conversion by \citet[][]{Zahnle2014} suggests that disequilibrium chemistry may delay the arrival of atmospheric $\rm CH_4$ in low gravity atmospheres to lower $T_{\rm eff}$ than current chemical disequilibrium models predict. This implies that a fresh set of atmospheric models accounting for this new understanding of the chemistry may be required before fully satisfactory models of young giant planets are in hand.

\section{Acknowledgments}
The Gemini Observatory is operated by AURA on behalf of the Gemini partnership. We acknowledge financial support of Gemini, NSF, and NASA. Portions of this work were performed under the auspices of the U.S. Department of Energy by Lawrence Livermore National Laboratory under Contract DE-AC52-07NA27344. We also thank the large team of scientists, engineers, technicians, and others who labored to make GPI a reality.


\begin{thebibliography}{}
\expandafter\ifx\csname natexlab\endcsname\relax\def\natexlab#1{#1}\fi

\bibitem[{{Ackerman} \& {Marley}(2001)}]{Ackerman01a}
{Ackerman}, A.~S., \& {Marley}, M.~S. 2001, \apj, 556, 872

\bibitem[{{Asplund} {et~al.}(2009){Asplund}, {Grevesse}, {Sauval}, \&
  {Scott}}]{asplund2009}
{Asplund}, M., {Grevesse}, N., {Sauval}, A.~J., \& {Scott}, P. 2009, \araa, 47,
  481

\bibitem[{{Baraffe} {et~al.}(2008){Baraffe}, {Chabrier}, \& {Barman}}]{bcb08}
{Baraffe}, I., {Chabrier}, G., \& {Barman}, T. 2008, \aap, 482, 315

\bibitem[{{Barman} {et~al.}(2011{\natexlab{a}}){Barman}, {Macintosh},
  {Konopacky}, \& {Marois}}]{Barman11a}
{Barman}, T.~S., {Macintosh}, B., {Konopacky}, Q.~M., \& {Marois}, C.
  2011{\natexlab{a}}, \apj, 733, 65

\bibitem[{{Barman} {et~al.}(2011{\natexlab{b}}){Barman}, {Macintosh},
  {Konopacky}, \& {Marois}}]{Barman11b}
---. 2011{\natexlab{b}}, \apjl, 735, L39

\bibitem[{{Beuzit} {et~al.}(2008){Beuzit}, {Feldt}, {Dohlen}, {Mouillet},
  {Puget}, {Wildi}, {Abe}, {Antichi}, {Baruffolo}, {Baudoz}, {Boccaletti},
  {Carbillet}, {Charton}, {Claudi}, {Downing}, {Fabron}, {Feautrier},
  {Fedrigo}, {Fusco}, {Gach}, {Gratton}, {Henning}, {Hubin}, {Joos}, {Kasper},
  {Langlois}, {Lenzen}, {Moutou}, {Pavlov}, {Petit}, {Pragt}, {Rabou}, {Rigal},
  {Roelfsema}, {Rousset}, {Saisse}, {Schmid}, {Stadler}, {Thalmann}, {Turatto},
  {Udry}, {Vakili}, \& {Waters}}]{Beuzit08a}
{Beuzit}, J.-L., {Feldt}, M., {Dohlen}, K., {et~al.} 2008, in Society of
  Photo-Optical Instrumentation Engineers (SPIE) Conference Series, Vol. 7014,
  Society of Photo-Optical Instrumentation Engineers (SPIE) Conference Series

\bibitem[{{Bowler} {et~al.}(2010){Bowler}, {Liu}, {Dupuy}, \&
  {Cushing}}]{Bowler10a}
{Bowler}, B.~P., {Liu}, M.~C., {Dupuy}, T.~J., \& {Cushing}, M.~C. 2010, \apj,
  723, 850

\bibitem[{{Currie} {et~al.}(2011){Currie}, {Burrows}, {Itoh}, {Matsumura},
  {Fukagawa}, {Apai}, {Madhusudhan}, {Hinz}, {Rodigas}, {Kasper}, {Pyo}, \&
  {Ogino}}]{Currie11a}
{Currie}, T., {Burrows}, A., {Itoh}, Y., {et~al.} 2011, \apj, 729, 128

\bibitem[{{Cushing} {et~al.}(2005){Cushing}, {Rayner}, \& {Vacca}}]{Cushing05a}
{Cushing}, M.~C., {Rayner}, J.~T., \& {Vacca}, W.~D. 2005, \apj, 623, 1115

\bibitem[{{Fortney} {et~al.}(2007){Fortney}, {Marley}, \& {Barnes}}]{FMB07}
{Fortney}, J.~J., {Marley}, M.~S., \& {Barnes}, J.~W. 2007, \apj, 659, 1661

\bibitem[{{Galicher} {et~al.}(2011){Galicher}, {Marois}, {Macintosh}, {Barman},
  \& {Konopacky}}]{Galicher11a}
{Galicher}, R., {Marois}, C., {Macintosh}, B., {Barman}, T., \& {Konopacky}, Q.
  2011, \apjl, 739, L41

\bibitem[{{Hinkley} {et~al.}(2011){Hinkley}, {Oppenheimer}, {Zimmerman},
  {Brenner}, {Parry}, {Crepp}, {Vasisht}, {Ligon}, {King}, {Soummer},
  {Sivaramakrishnan}, {Beichman}, {Shao}, {Roberts}, {Bouchez}, {Dekany},
  {Pueyo}, {Roberts}, {Lockhart}, {Zhai}, {Shelton}, \& {Burruss}}]{Hinkley11a}
{Hinkley}, S., {Oppenheimer}, B.~R., {Zimmerman}, N., {et~al.} 2011, \pasp,
  123, 74

\bibitem[{{Hinz} {et~al.}(2010){Hinz}, {Rodigas}, {Kenworthy}, {Sivanandam},
  {Heinze}, {Mamajek}, \& {Meyer}}]{Hinz10a}
{Hinz}, P.~M., {Rodigas}, T.~J., {Kenworthy}, M.~A., {et~al.} 2010, \apj, 716,
  417

\bibitem[{{Janson} {et~al.}(2010){Janson}, {Bergfors}, {Goto}, {Brandner}, \&
  {Lafreni{\`e}re}}]{Janson10a}
{Janson}, M., {Bergfors}, C., {Goto}, M., {Brandner}, W., \& {Lafreni{\`e}re},
  D. 2010, \apjl, 710, L35

\bibitem[{{Konopacky} {et~al.}(2013){Konopacky}, {Barman}, {Macintosh}, \&
  {Marois}}]{Konopacky13a}
{Konopacky}, Q.~M., {Barman}, T.~S., {Macintosh}, B.~A., \& {Marois}, C. 2013,
  Science, 339, 1398

\bibitem[{Larkin {et~al.}(2006)Larkin, Barczys, Krabbe, Adkins, Aliado, Amico,
  Brims, Campbell, Canfield, Gasaway, Honey, Iserlohe, Johnson, Kress,
  LaFreniere, Lyke, Magnone, Magnone, McElwain, Moon, Quirrenbach, Skulason,
  Song, Spencer, Weiss, \& Wright}]{Larkin06a}
Larkin, J., Barczys, M., Krabbe, A., {et~al.} 2006, OSIRIS: a diffraction
  limited integral field spectrograph for Keck, doi:10.1117/12.672061

\bibitem[{Larkin {et~al.}(2014)Larkin, Chilcote, Aliado, Bauman, Brims,
  Canfield, Cardwell, Dillon, Doyon, Dunn, Fitzgerald, Graham, Goodsell,
  Hartung, Hibon, Ingraham, Johnson, Kress, Konopacky, Macintosh, Magnone,
  Maire, McLean, Palmer, Perrin, Quiroz, Rantakyrö, Sadakuni, Saddlemyer,
  Serio, Thibault, Thomas, Vallee, \& Weiss}]{Larkin14a}
Larkin, J.~E., Chilcote, J.~K., Aliado, T., {et~al.} 2014, in Society of
  Photo-Optical Instrumentation Engineers (SPIE) Conference Series, Vol. 9147,
  Society of Photo-Optical Instrumentation Engineers (SPIE) Conference Series

\bibitem[{{Lee} {et~al.}(2013){Lee}, {Heng}, \& {Irwin}}]{Lee13a}
{Lee}, J.-M., {Heng}, K., \& {Irwin}, P.~G.~J. 2013, \apj, 778, 97

\bibitem[{{Lodders}(2003)}]{lodders2003}
{Lodders}, K. 2003, \apj, 591, 1220

\bibitem[{Macintosh {et~al.}(2014)Macintosh, Graham, Ingraham, Konopacky,
  Marois, Perrin, Poyneer, Bauman, Barman, Burrows, Cardwell, Chilcote,
  De~Rosa, Dillon, Doyon, Dunn, Erikson, Fitzgerald, Gavel, Goodsell, Hartung,
  Hibon, Kalas, Larkin, Maire, Marchis, Marley, McBride, Millar-Blanchaer,
  Morzinski, Norton, Oppenheimer, Palmer, Patience, Pueyo, Rantakyro, Sadakuni,
  Saddlemyer, Savransky, Serio, Soummer, Sivaramakrishnan, Song, Thomas,
  Wallace, Wiktorowicz, \& Wolff}]{Macintosh14a}
Macintosh, B., Graham, J.~R., Ingraham, P., {et~al.} 2014, Proceedings of the
  National Academy of Sciences,
  http://www.pnas.org/content/early/2014/05/08/1304215111.full.pdf+html

\bibitem[{Macintosh {et~al.}(2014b)Macintosh, Graham, Ingraham, Konopacky,
  Marois, Perrin, Poyneer, Bauman, Barman, Burrows, Cardwell, Chilcote,
  De~Rosa, Dillon, Doyon, Dunn, Erikson, Fitzgerald, Gavel, Goodsell, Hartung,
  Hibon, Kalas, Larkin, Maire, Marchis, Marley, McBride, Millar-Blanchaer,
  Morzinski, Norton, Oppenheimer, Palmer, Patience, Pueyo, Rantakyro, Sadakuni,
  Saddlemyer, Savransky, Serio, Soummer, Sivaramakrishnan, Song, Thomas,
  Wallace, Wiktorowicz, \& Wolff}]{Macintosh14b}
Macintosh, B., Graham, J.~R., Ingraham, P., {et~al.} 2014b, in Society of
  Photo-Optical Instrumentation Engineers (SPIE) Conference Series, Vol. 9148,
  Society of Photo-Optical Instrumentation Engineers (SPIE) Conference Series

\bibitem[{{Madhusudhan} {et~al.}(2011){Madhusudhan}, {Burrows}, \&
  {Currie}}]{Madhusudan11a}
{Madhusudhan}, N., {Burrows}, A., \& {Currie}, T. 2011, \apj, 737, 34

\bibitem[{{Marley} {et~al.}(2012){Marley}, {Saumon}, {Cushing}, {Ackerman},
  {Fortney}, \& {Freedman}}]{Marley12a}
{Marley}, M.~S., {Saumon}, D., {Cushing}, M., {et~al.} 2012, \apj, 754, 135

\bibitem[{{Marley} {et~al.}(2010){Marley}, {Saumon}, \&
  {Goldblatt}}]{Marley10a}
{Marley}, M.~S., {Saumon}, D., \& {Goldblatt}, C. 2010, \apjl, 723, L117

\bibitem[{{Marois} {et~al.}(2014){Marois}, {Correia}, {V{\'e}ran}, \&
  {Currie}}]{Marois14a}
{Marois}, C., {Correia}, C., {V{\'e}ran}, J.-P., \& {Currie}, T. 2014, in IAU
  Symposium, Vol. 299, IAU Symposium, ed. M.~{Booth}, B.~C. {Matthews}, \&
  J.~R. {Graham}, 48--49

\bibitem[{{Marois} {et~al.}(2008){Marois}, {Macintosh}, {Barman}, {Zuckerman},
  {Song}, {Patience}, {Lafreni{\`e}re}, \& {Doyon}}]{marois08a}
{Marois}, C., {Macintosh}, B., {Barman}, T., {et~al.} 2008, Science, 322, 1348

\bibitem[{{Marois} {et~al.}(2010){Marois}, {Zuckerman}, {Konopacky},
  {Macintosh}, \& {Barman}}]{marois10a}
{Marois}, C., {Zuckerman}, B., {Konopacky}, Q.~M., {Macintosh}, B., \&
  {Barman}, T. 2010, \nat, 468, 1080

\bibitem[{{Martinache} \& {Guyon}(2009)}]{Martinache09a}
{Martinache}, F., \& {Guyon}, O. 2009, in Society of Photo-Optical
  Instrumentation Engineers (SPIE) Conference Series, Vol. 7440, Society of
  Photo-Optical Instrumentation Engineers (SPIE) Conference Series

\bibitem[{{Oppenheimer} {et~al.}(2013){Oppenheimer}, {Baranec}, {Beichman},
  {Brenner}, {Burruss}, {Cady}, {Crepp}, {Dekany}, {Fergus}, {Hale},
  {Hillenbrand}, {Hinkley}, {Hogg}, {King}, {Ligon}, {Lockhart}, {Nilsson},
  {Parry}, {Pueyo}, {Rice}, {Roberts}, {Roberts}, {Shao}, {Sivaramakrishnan},
  {Soummer}, {Truong}, {Vasisht}, {Veicht}, {Vescelus}, {Wallace}, {Zhai}, \&
  {Zimmerman}}]{Oppenheimer13a}
{Oppenheimer}, B.~R., {Baranec}, C., {Beichman}, C., {et~al.} 2013, \apj, 768,
  24

\bibitem[{Perrin {et~al.}(2014)Perrin, Maire, Ingraham, Savransky,
  Millar-Blanchaer, Wolff, Ruffio, Wang, Draper, Sadakuni, Marois, Fitzgerald,
  Macintosh, Graham, Doyon, Larkin, Chilcote, Goodsell, Palmer, Labrie,
  Beaulieau, Rosa, Greenbaum, Hartung, Hibon, Konopacky, Lafreniere, Lavigne,
  Marchis, Patience, Pueyo, Soummer, Thomas, Ward-Duong, \&
  Wiktorowicz}]{Perrin14a}
Perrin, M., Maire, J., Ingraham, P.~J., {et~al.} 2014, in Society of
  Photo-Optical Instrumentation Engineers (SPIE) Conference Series, Vol. 9147,
  Society of Photo-Optical Instrumentation Engineers (SPIE) Conference Series

\bibitem[{{Saumon} \& {Marley}(2008)}]{sm08}
{Saumon}, D., \& {Marley}, M.~S. 2008, \apj, 689, 1327

\bibitem[{{Skemer} {et~al.}(2012){Skemer}, {Hinz}, {Esposito}, {Burrows},
  {Leisenring}, {Skrutskie}, {Desidera}, {Mesa}, {Arcidiacono}, {Mannucci},
  {Rodigas}, {Close}, {McCarthy}, {Kulesa}, {Agapito}, {Apai}, {Argomedo},
  {Bailey}, {Boutsia}, {Briguglio}, {Brusa}, {Busoni}, {Claudi}, {Eisner},
  {Fini}, {Follette}, {Garnavich}, {Gratton}, {Guerra}, {Hill}, {Hoffmann},
  {Jones}, {Krejny}, {Males}, {Masciadri}, {Meyer}, {Miller}, {Morzinski},
  {Nelson}, {Pinna}, {Puglisi}, {Quanz}, {Quiros-Pacheco}, {Riccardi},
  {Stefanini}, {Vaitheeswaran}, {Wilson}, \& {Xompero}}]{Skemer12a}
{Skemer}, A.~J., {Hinz}, P.~M., {Esposito}, S., {et~al.} 2012, \apj, 753, 14

\bibitem[{Skemer {et~al.}(2014)Skemer, Marley, Hinz, Morzinski, Skrutskie,
  Leisenring, Close, Saumon, Bailey, Briguglio, Defrere, Esposito, Follette,
  Hill, Males, Puglisi, Rodigas, \& Xompero}]{Skemer14a}
Skemer, A.~J., Marley, M.~S., Hinz, P.~M., {et~al.} 2014, ApJ in press,
  arXiv:1311.2085

\bibitem[{{Stephens} {et~al.}(2009){Stephens}, {Leggett}, {Cushing}, {Marley},
  {Saumon}, {Geballe}, {Golimowski}, {Fan}, \& {Noll}}]{Stephens09a}
{Stephens}, D.~C., {Leggett}, S.~K., {Cushing}, M.~C., {et~al.} 2009, \apj,
  702, 154

\bibitem[{Wang {et~al.}(2014)Wang, Rajan, Graham, Savransky, Ingraham,
  Ward-Duong, Patience, Rosa, Bulger, Sivaramakrishnan, Perrin, Thomas,
  Sadakuni, Greenbaum, Pueyo, Marois, Oppenheimer, Kalas, Cardwell, Goodsell,
  Hibon, \& Rantakyr\"o}]{Wang14a}
Wang, J.~J., Rajan, A., Graham, J.~R., {et~al.} 2014, in

\bibitem[{{Zahnle} \& {Marley}(2014)}]{Zahnle2014}
{Zahnle}, K.~J., \& {Marley}, M.~S. 2014, ArXiv e-prints, arXiv:1408.6283

\end{thebibliography}
\end{document}